\documentclass[a4paper]{ESASPCS13Style}
\usepackage{epsfig}

\begin{document}

\title{Cool Stars 13: Spectral Classification Beyond M }

\author{S.\,K. Leggett\inst{1} \and 
F. Allard \inst{2} \and
A.\,J. Burgasser \inst{3} \and
H.\,R.\,A. Jones \inst{4} \and
M.\,S. Marley  \inst{5} \and
T. Tsuji \inst{6}} 
\institute{Joint Astronomy Centre Hawaii
\and Centre de Recherche Astronomique de Lyon
\and UCLA Department Physics and Astronomy
\and University of Hertfordshire
\and NASA Ames Research Center
\and Institute of Astronomy University of Tokyo }

\maketitle 

\begin{abstract}

Significant populations of field L and T dwarfs are now known, and we anticipate the discovery of even cooler dwarfs by Spitzer and
ground--based infrared surveys. However, as the number of known L and T dwarfs increases so does the range in their observational properties, and difficulties have arisen in interpreting the observations. Although
modellers have made significant advances, the complexity of the very low 
temperature, high pressure, photospheres means that problems remain such 
as the treatment of grain condensation as well as incomplete and non-equilibrium molecular chemistry. Also, there are several parameters which control the observed spectral energy distribution --- effective temperature, grain sedimentation efficiency, metallicity and gravity --- and their effects are not well understood.  In this paper, based on a splinter session, we discuss classification schemes for L and T dwarfs, their dependency on wavelength, and the effects of the parameters $T_{\rm eff}$, $f_{\rm sed}$, [$m/H$] and 
log$g$ on optical and infrared spectra. We will also discuss the various hypotheses that have been presented for the transition from the dusty L types to the clear atmosphere T types.  We conclude with a brief discussion
of the spectral class beyond T.  Authors of each Section are identified by their initials.

\end{abstract}

\section{Current Spectral Classification Schemes 
({\it SKL})   }

Both optical and infrared classification schemes exist for L dwarfs.
In 1999, Kirkpatrick et al. and Mart\'{\i}n et al. published optical schemes
that use the strength of various features (VO 7912~$\AA$, Rb 7948~$\AA$, TiO 8432~$\AA$, Cs 8521~$\AA$, CrH 8611~$\AA$) and pseudo--continuum slopes (the color--d 9775/7450 or PC3 8250/7560 flux ratios) to classify L dwarfs.
The Mart\'{\i}n et al. scheme terminates at L6 and the Kirkpatrick et al. at L8; types derived from the two schemes agree well except for later types, where the Mart\'{\i}n scheme produces earlier types than Kirkpatrick.  A large number of L dwarfs have been identified by the  2 Micron All Sky Survey (2MASS) and classified on the Kirkpatrick scheme, which has therefore become the commonly used L dwarf classification scheme.

Geballe et al. (2002) published an infrared classification scheme
for L dwarfs that used a modified color--d index for types L0 to L6,
the strength of the 1.5~$\mu$m wing of the H$_2$O absorption band for L0 to
L9, and the shape of the 2.2~$\mu$m flux peak (as measured by a
CH$_4$ index) for types L3 to L9.  This scheme was pinned to the
Kirkpatrick et al. and Mart\'{\i}n et al. schemes for earlier L types and agrees reasonably well with the Kirkpatrick optical scheme, but differences of up to two subclasses do exist.

\begin{figure}[!h]
\begin{center}
 \includegraphics[height=.35\textheight,angle=-90]{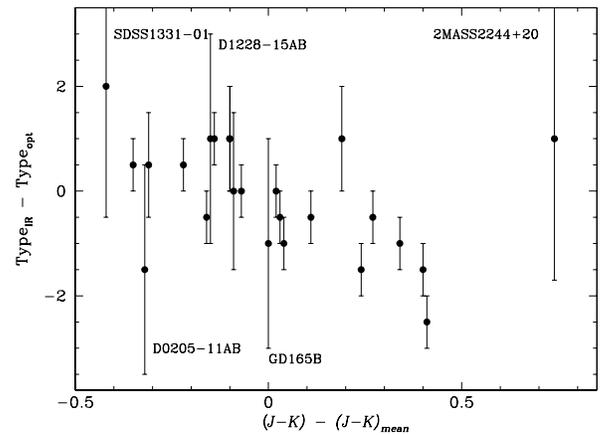}
\end{center}
  \caption{Difference between infrared and optical L type for 
a sample of L2.5 to L8 dwarfs.}
\end{figure}

We can understand these optical to infrared differences in terms of the effect of the silicate cloud decks that exist in the photospheres of mid-- to late--type L dwarfs.  For these objects, the regions of the photosphere probed by the different indices are a strong function of wavelength (see for example Figure 7 of Ackerman \& Marley 2001).  Where the atmosphere is opaque, e.g. in the far--red and $K$ bands, the flux emerges from regions above the cloud decks which are less sensitive to the cloud optical depth.  However the clear $Z$ and $J$ regions are very sensitive to the clouds.  Hence the 1--2~$\mu$m infrared indices may be more indicative of cloud optical depth than $T_{\rm eff}$.  Also, we would expect the difference between optical and infrared type to be a function of color, as the redder, dustier, dwarfs will suffer more veiling and hence the infrared type will be earlier than the optical type.  Such a trend is seen in Figure 1 (excluding the extremely red L dwarf 2MASS2244+20 for which the optical type also seems to be affected by the clouds).  The more unusually blue or red L dwarfs also show a scatter between the infrared indices, as indicated by the error bars in Figure 1 (see discussion in Knapp et al. 2004).  The optical Kirkpatrick scheme terminates at L8 whereas the infrared Geballe scheme requires an L9
type.  Of six infrared L9---T0 types with both optical and infrared classifications, four are optical L8s and two are L5s; one of the latter is very blue in the infrared and would be expected to have a very early L optical type.  

Optical and infrared classification schemes also exist for the T dwarfs, however the optical scheme is recognised to be less accurate than the infrared schemes, and the latter are commonly used.  There is little flux in the optical for the T dwarfs, making classification at these wavelengths difficult, and very few early T dwarfs have good signal to noise optical spectra, making standardisation less accurate.  The optical scheme is published by Burgasser et al. (2003) and uses the Cs~I 8521~$\AA$, CrH 
8611~$\AA$, H$_2$O 9250~$\AA$ and FeH 9896~$\AA$ features (the last for types
T5 and later only), as well as the color--e 9190/8450 flux ratio (for T2 and earlier).  Infrared schemes have been published by Burgasser et al. (2002)
and Geballe et al. (2002).  Both use the strengths of the 1.2~$\mu$m and 
1.5~$\mu$m H$_2$O bands, and the 1.6~$\mu$m and 2.2~$\mu$m CH$_4$ bands; Burgasser also uses an additional 1.3 $\mu$m CH$_4$ band and $H/J$, $K/J$, 2.11/2.07 flux ratios.  The infrared schemes agree very well, typically to within half a subclass.  The schemes are very similar and a combined scheme is in preparation (Burgasser et al.).

\begin{figure}[!h]
\begin{center}
 \includegraphics[height=.35\textheight,angle=-90]{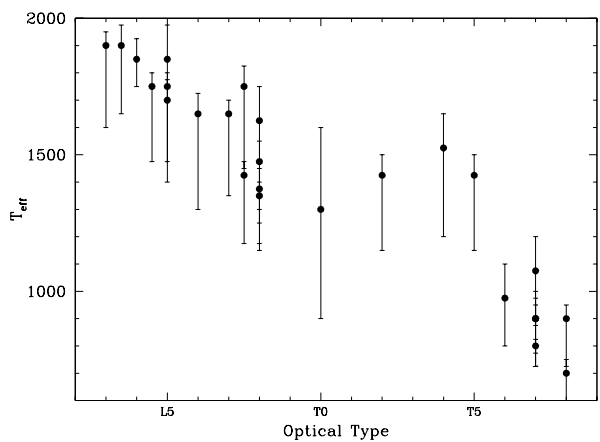}
\end{center}
  \caption{Effective temperatures from Golimoski et al. (2004) plotted as a function of optical L and T type.}
\end{figure}

An important issue is whether current classification schemes can be used as indicators of atmospheric parameters, such as $T_{\rm eff}$.  Golimoski et al. (2004) have derived $T_{\rm eff}$ by integrating flux calibrated spectra for dwarfs with measured trigonometric parallaxes.  Figure 6 of Golimoski et al. plots $T_{\rm eff}$ as a function of infrared L and T type.  Figure 2 reproduces this using optical type for both the Ls and Ts.  Whichever scheme is used, we find that $T_{\rm eff}$ is approximately constant at 1450 K for L7 to T4 types.  That is, the observed spectral changes are not due to decreasing temperature, but must be explained by some other means, such as clearing of the cloud decks (see discussion in \S 3).  Note that this implies that if the optical type is a pure indicator of $T_{\rm eff}$ then dwarfs with infrared type L7 to T4 should all have optical type L8.  The other important atmospheric parameters, metallicity and gravity, are discussed in the next Section.

\subsection{Discussion ({\it All})}

$Q$: Could the derived constant $T_{\rm eff}$ be an error caused somehow by the breakup of the dust clouds? \break
$A$: The derivation is fairly secure as it relies on summing observed flux and well understood structural models of brown dwarf radii.  [Elsewhere in these proceedings Cushing shows that the Spitzer mid--infrared results confirm the Golimoski et al. bolometric luminosities.] \break
$Q$: Can you expand on the the term ``veiling''?\break
$A$: This may be misleading.  The spectra of a dusty L dwarf is reddened and
has weakened molecular absorption bands similar to a veiling effect.  What is actually happening is that the atmosphere is heated by the dust and so the bands are formed in hotter regions and are therefore weaker.  See further discussion by Allard in this session.

\section{Other Dimensions to Classification}

\subsection{Observed Indicators of [$m/H$], log$g$ 
({\it AB})}

Overlapping and blanketing absorption bands from the various chemically active
atomic and molecular species present in the atmospheres of cool L and T dwarfs imply
emergent spectra that are particularly
sensitive to metallicity and gravity effects.
Gravity diagnostics have been investigated for
low--mass brown dwarf candidates in young star--forming regions, which
can have surface gravities 10---100 times lower than those of  field dwarfs.
Building from a substantial body of work on
young M--type brown dwarf candidates and giant stars
(e.g. Steele \& Jameson 1995; Mart{\'{i}}n, Rebolo \& Zapatero Osorio 1996;
Luhman \& Rieke 1998; Gorlova et al.\ 2003; Slesnick, Hillenbrand \& Carpenter 2004),
investigators are now searching for gravity diagnostics in young L dwarf
spectra.  Lucas et al.\ (2001) classify a number of
faint Trapezium brown dwarf candidates as L--type dwarfs
based on the strength of H$_2$O bands, calibrated against field dwarf spectra.
However, the spectral morphologies of these objects are
quite different from those of equivalent field dwarfs, with triangular $H$--band
peaks and weakened Na~I and CO absorption features.  Near--infrared spectra
of low--mass brown dwarf candidates in $\sigma$ Orionis obtained by
Mart{\'{i}}n et al.\ (2001) are similarly classified as L dwarfs
but do not show the unusual $H$--band morphologies.  A comparison to optical
spectral types obtained by Barrado y Navascu{\'{e}}s et al.\ (2001) for two
of the $\sigma$ Orionis sources suggests that
near--infrared types for these low--gravity L dwarfs may be overestimated.
McGovern et al.\ (2004) examined optical and near--infrared spectra
for more reliable young cluster brown dwarf candidates and the $\sim$20---300 Myr
companion L dwarf G 196-3B (Rebolo et al.\ 1998; Figure 3),
finding that many of the spectral features indicative of
low--gravity M dwarfs and giants --- enhanced metal oxides and H$_2$O, weak alkali lines, and weak metal
hydrides --- are also gravity diagnostics in the L dwarf regime.  McGovern et al.\ used these
gravity diagnostics to rule out cluster
membership for at least one of the $\sigma$ Orionis candidates in the Mart{\'{i}}n et al.\ (2001)
study.  Future applications of this technique may ultimately enable more reliable measures of
the substellar mass function in star forming regions.
A more rigorous study of L dwarf gravity diagnostics
is being conducted by J.~D. Kirkpatrick (priv. comm.).

\begin{figure}[ht]
  \begin{center}
    \epsfig{file=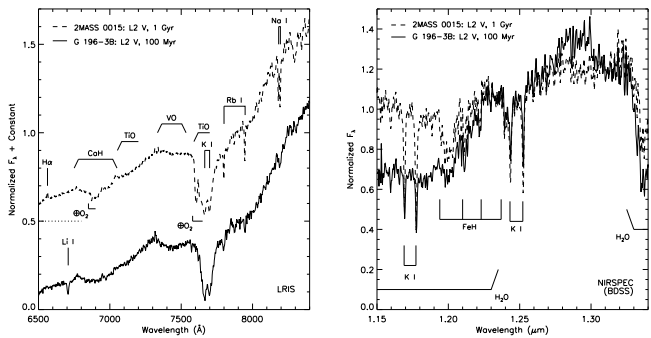, width=9cm}
  \end{center}
\caption{Low--gravity features in the 20---300 Myr
L2 brown dwarf G 196-3B compared to a field L2
dwarf 2MASS~0015+35.  At optical wavelengths, G 196-3B exhibits
weakened alkali lines (e.g. Rb I, Na I) while TiO bands are
somewhat enhanced. At near--infrared wavelengths, H$_2$O
bands are enhanced, while alkali lines and metal hydride bands
(e.g. FeH) are weaker.  (Data courtesy J.\ D.\ Kirkpatrick and I.\ S.\ McLean.)
\label{fig_adam1}}
\end{figure}

The identification of gravity features in young cluster T dwarfs has been
largely impeded by their intrinsic faintness, but some progress has been made
amongst higher--gravity, old and massive field brown dwarfs.
Two strongly gravity--sensitive
features are present in the spectra of these objects: the pressure--broadened K I and Na I
fundamental doublet lines
that dominate optical opacity (Burrows, Marley, \& Sharp 2000) and 
collision--induced
H$_2$ absorption centered around 2 $\mu$m (Saumon et al.\ 1994).  Both features
are enhanced in the spectra of higher gravity sources, while other strong molecular bands
(e.g. CH$_4$, H$_2$O) are more sensitive to $T_{\rm eff}$.  Variations of these features
amongst similarly classified field T dwarfs
has been shown by Burgasser et al.\ (2005 in prep., Figure 4);
while Knapp et al.\ (2004) have used $H-K$ color as a
diagnostic for surface gravity.  The near--infrared alkali lines behave in 
an opposite
manner in T dwarfs as compared to M and L dwarfs, with weaker lines present in the spectra of
higher gravity sources (Burgasser et al.\ 2002; Knapp et al.\ 2004).
Only one low--gravity young cluster T dwarf candidate has thus far been identified,
S Ori 70 (Zapatero Osorio et al.\ 2002; Mart{\'{i}}n \& Zapatero Osorio 2003), although
the cluster membership of this source remains controversial
(Burgasser et al.\ 2004).  A low--gravity field T dwarf may have
recently been identified in the SDSS survey by Knapp et al.\ (2004).

\begin{figure}[ht]
  \begin{center}
    \epsfig{file=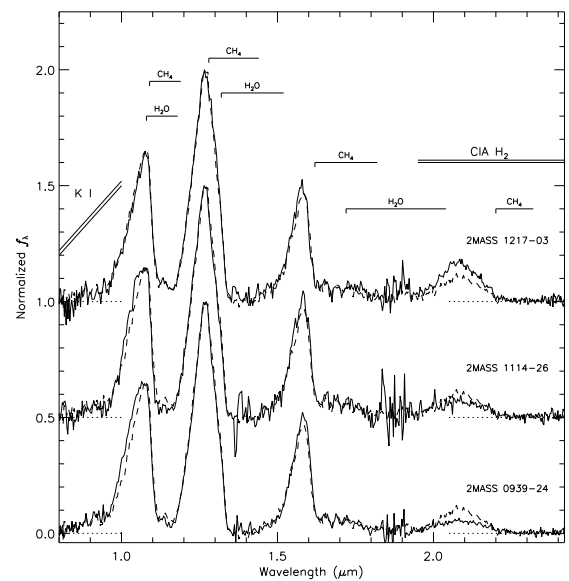, width=8cm}
  \end{center}
\caption{Low resolution near--infrared spectra of T dwarfs showing
variations in gravity--sensitive features.  All of the objects plotted
as solid lines are classified T7.5 based on the strength of H$_2$O and CH$_4$
bandstrengths and have similar spectral morphologies to the 2---5 Gyr
T7.5 companion brown dwarf Gliese 570D (Burgasser et al.\ 2000; dashed 
lines). Two exceptions are the 1 $\mu$m peak, shaped by the 
pressure--broadened
wings of K I, and the 2.2 $\mu$m peak, controlled by collision--induced
H$_2$ absorption. (Adapted from Burgasser et al.\ in prep.)
\label{fig_adam2}}
\end{figure}

\begin{figure}[!!b]
\begin{center}
\epsfig{file=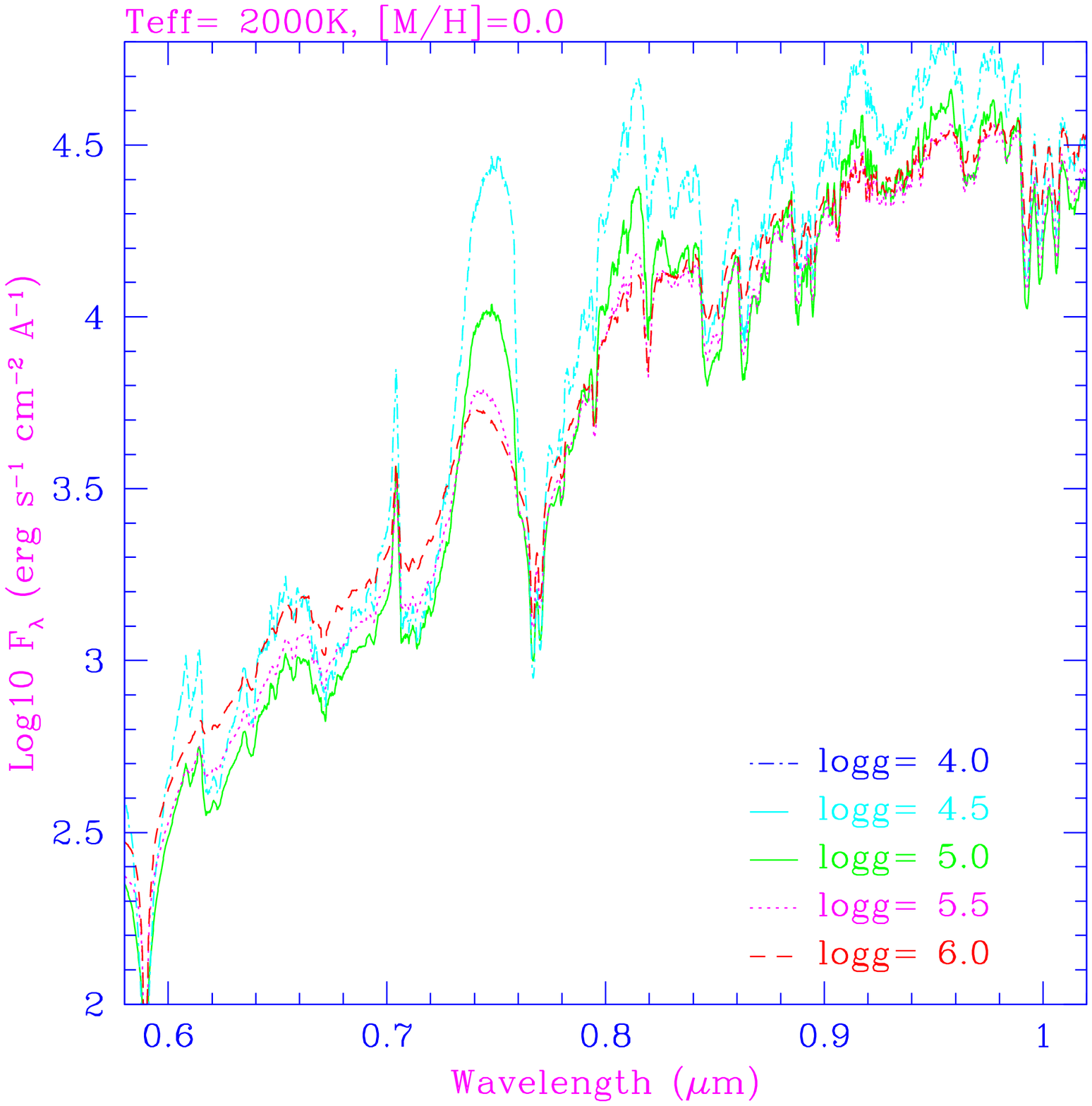,scale=0.4}
\epsfig{file=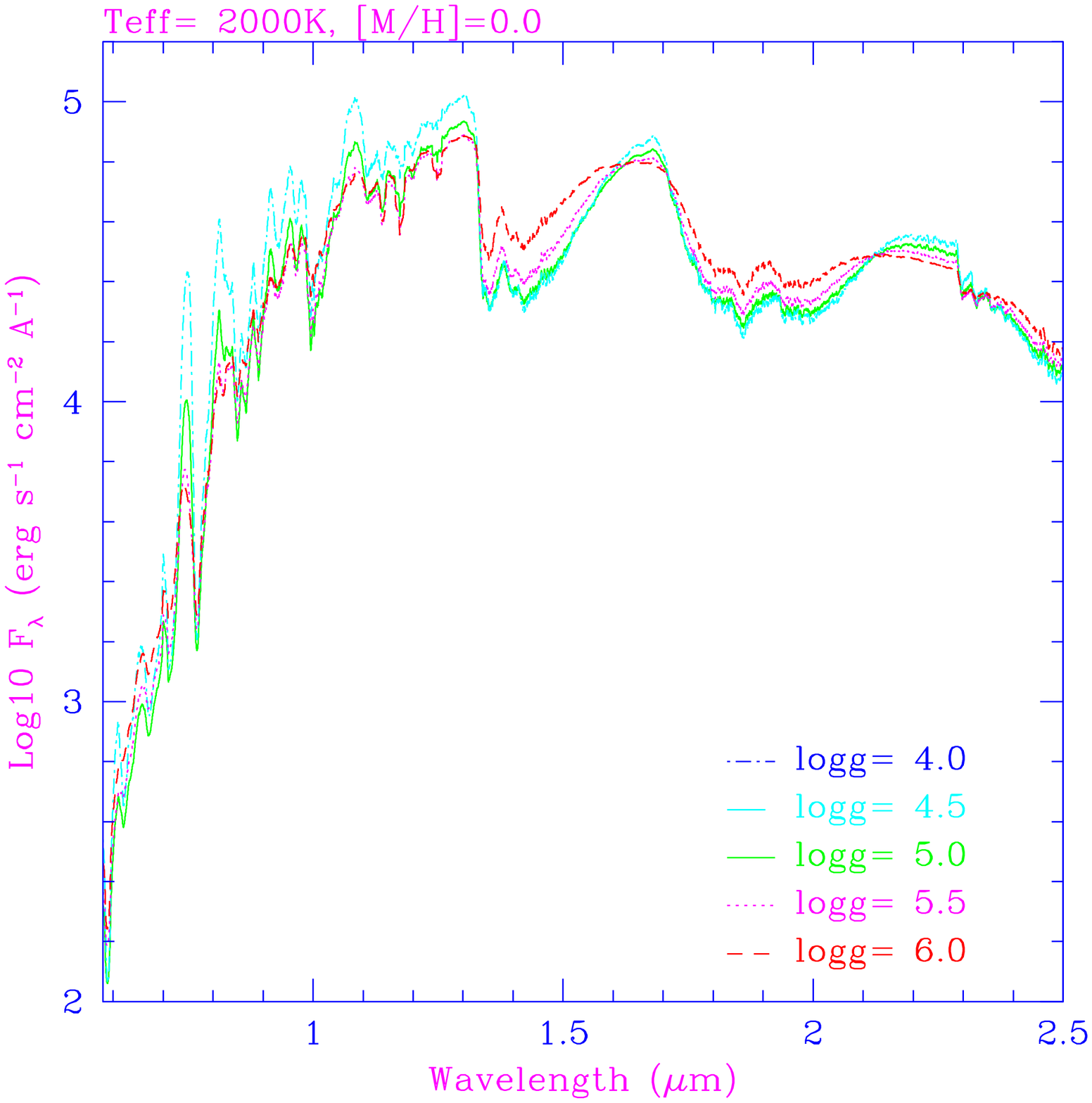,scale=0.4}
\caption{AMES-Dusty-2001 models for $T_{\rm eff}= 2000$~K, [$m/H$]$=$ 0.0 for log$g=$ 4.0 to 6.0  in the 
optical (top) and infrared (bottom). With increasing gravity, the dust scattering continuum increases, 
producing a backwarming that decreases the TiO and H$_2$O band strengths
(cf. Fig.3). }
\label{default}
\end{center}
\end{figure}


\begin{figure}[!!b]
\begin{center}
\epsfig{file=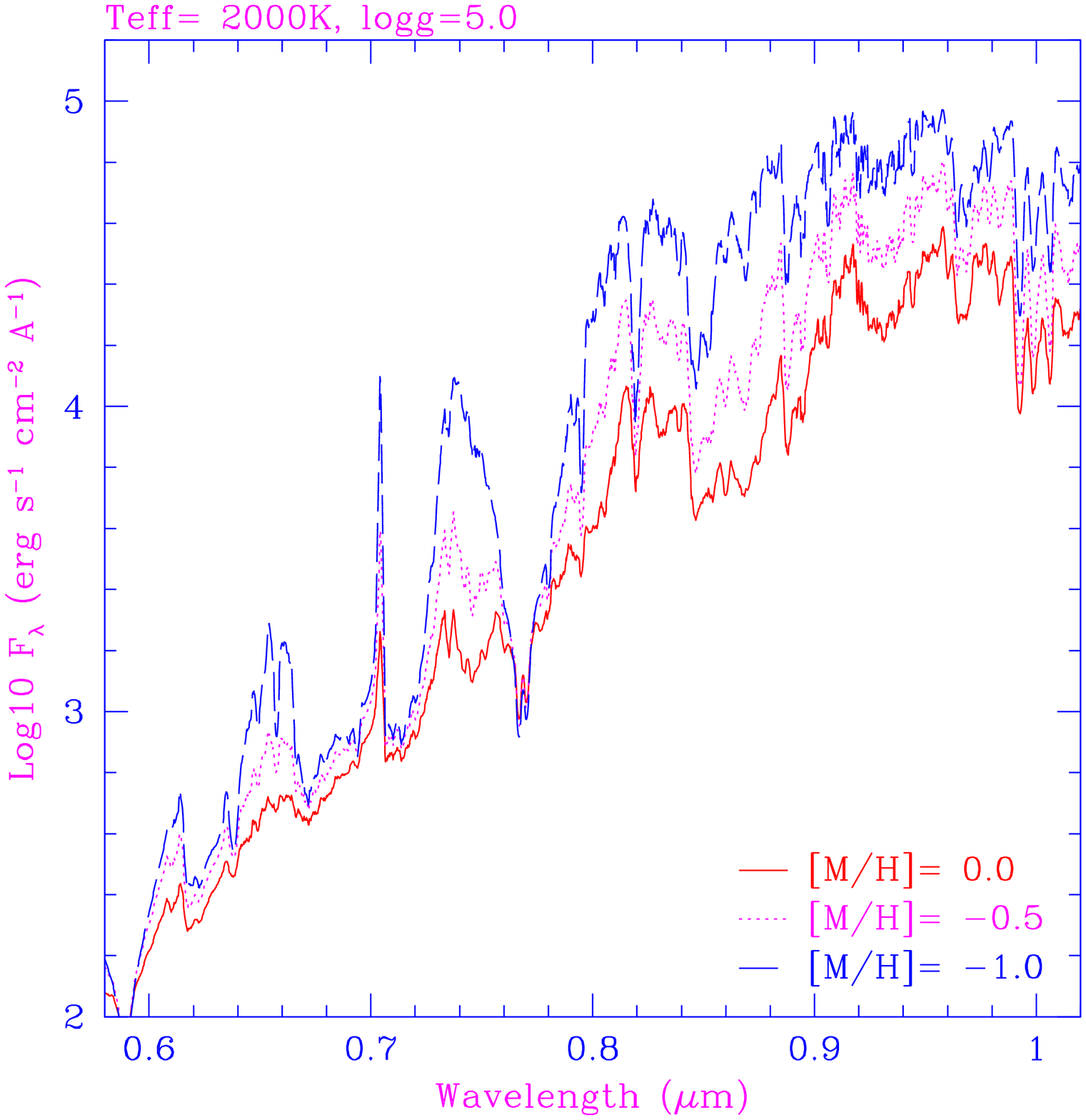,scale=0.4}
\epsfig{file=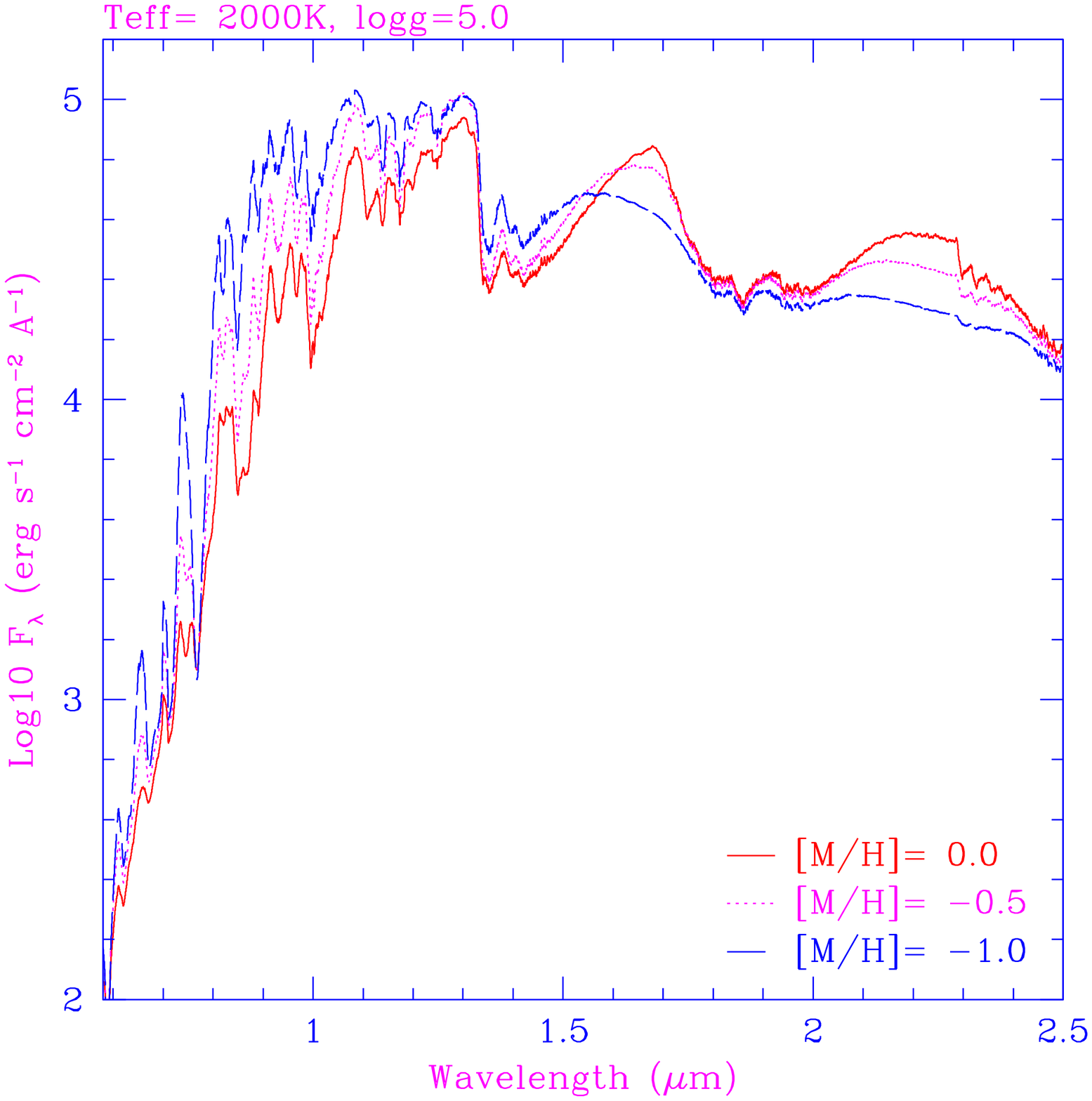,scale=0.4}
\caption{AMES-Dusty v2.7 models for $T_{\rm eff}= 2000$~K, log$g=$ 5.0 for [$m/H$]$=$ 0.0, -0.5, -1.0 in the 
optical (top) and infrared (bottom). With decreasing metallicity, the dust scattering, TiO and VO opacities 
decrease, and H$_2$ opacity increasingly depresses the $H$ and $K$ flux peaks. }
\label{default}
\end{center}
\end{figure}

\begin{figure}[!t]
\begin{center}
\epsfig{file=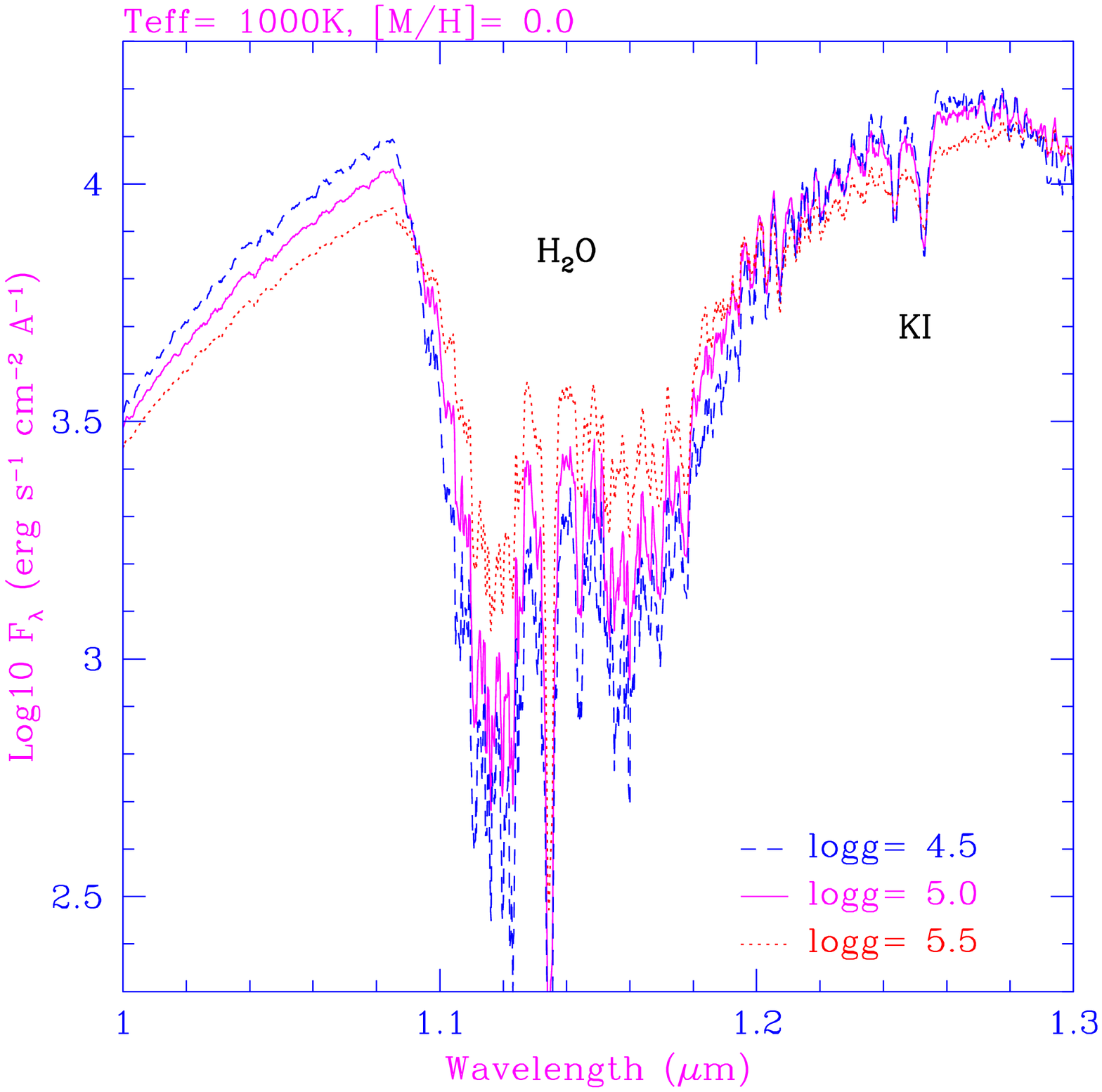,scale=0.4}
\epsfig{file=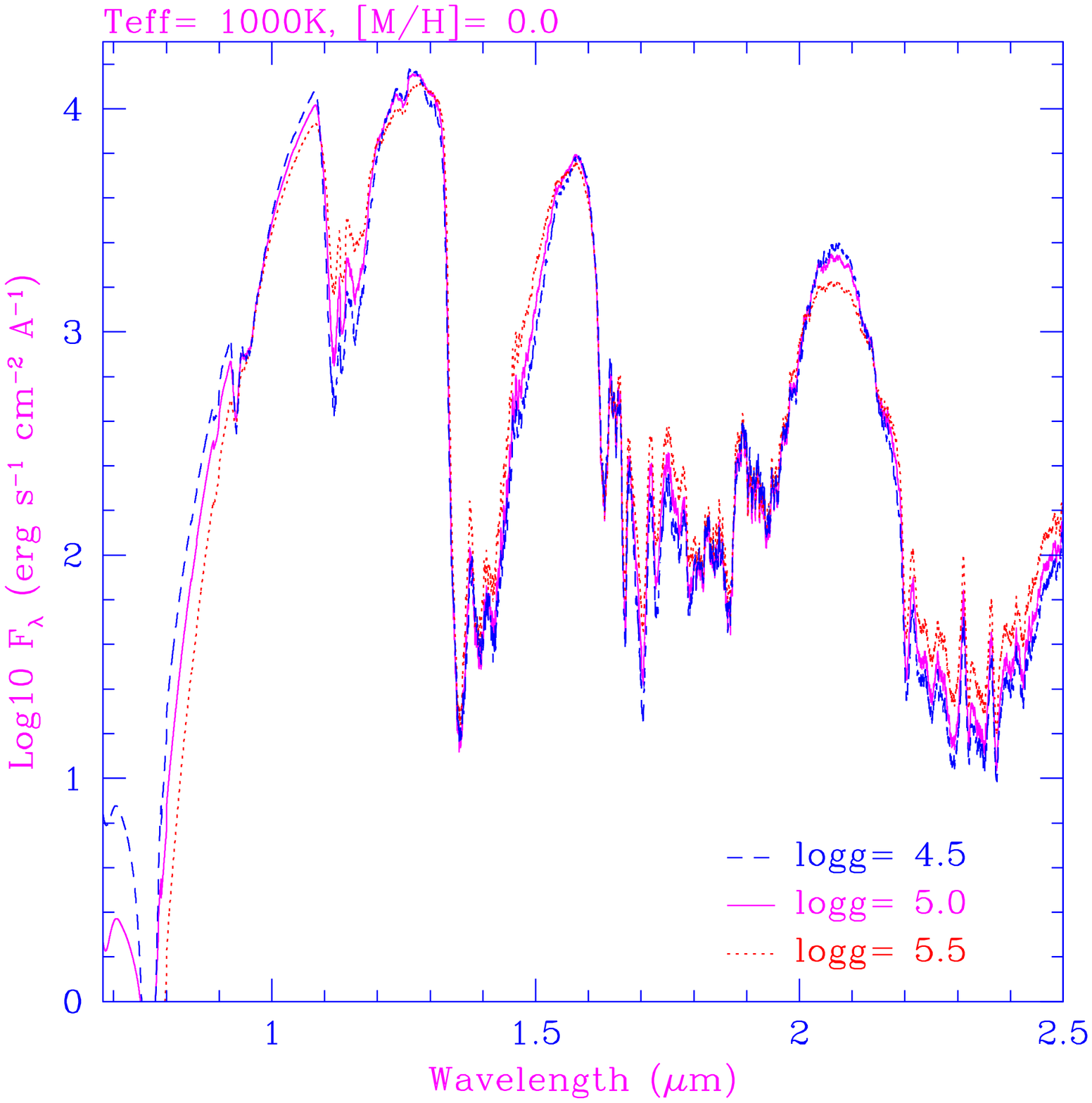,scale=0.4}
\caption{
AMES-Rainout models using detailed Na~ID and 0.77~$\mu$m K~I 
line profiles (Allard et al. 2003) for $T_{\rm eff}=1000$~K, 
[$m/H$]$=0.0$ are compared for log$g=$4.5, 5.0 and 5.5. With increasing 
gravity, the broadening of the K~I doublets  at 0.77 and 1.25~$\mu$m 
increases with  gas pressure.  The $K$--band flux peak is 
depressed by increasing H$_2$ opacity, and the strengths of the 1.14 
and 2.3 $\mu$m H$_2$O bands decrease (cf. Figure 4). 
}
\label{default}
\end{center}
\end{figure}

\begin{figure}[!ht]
\begin{center}
\epsfig{file=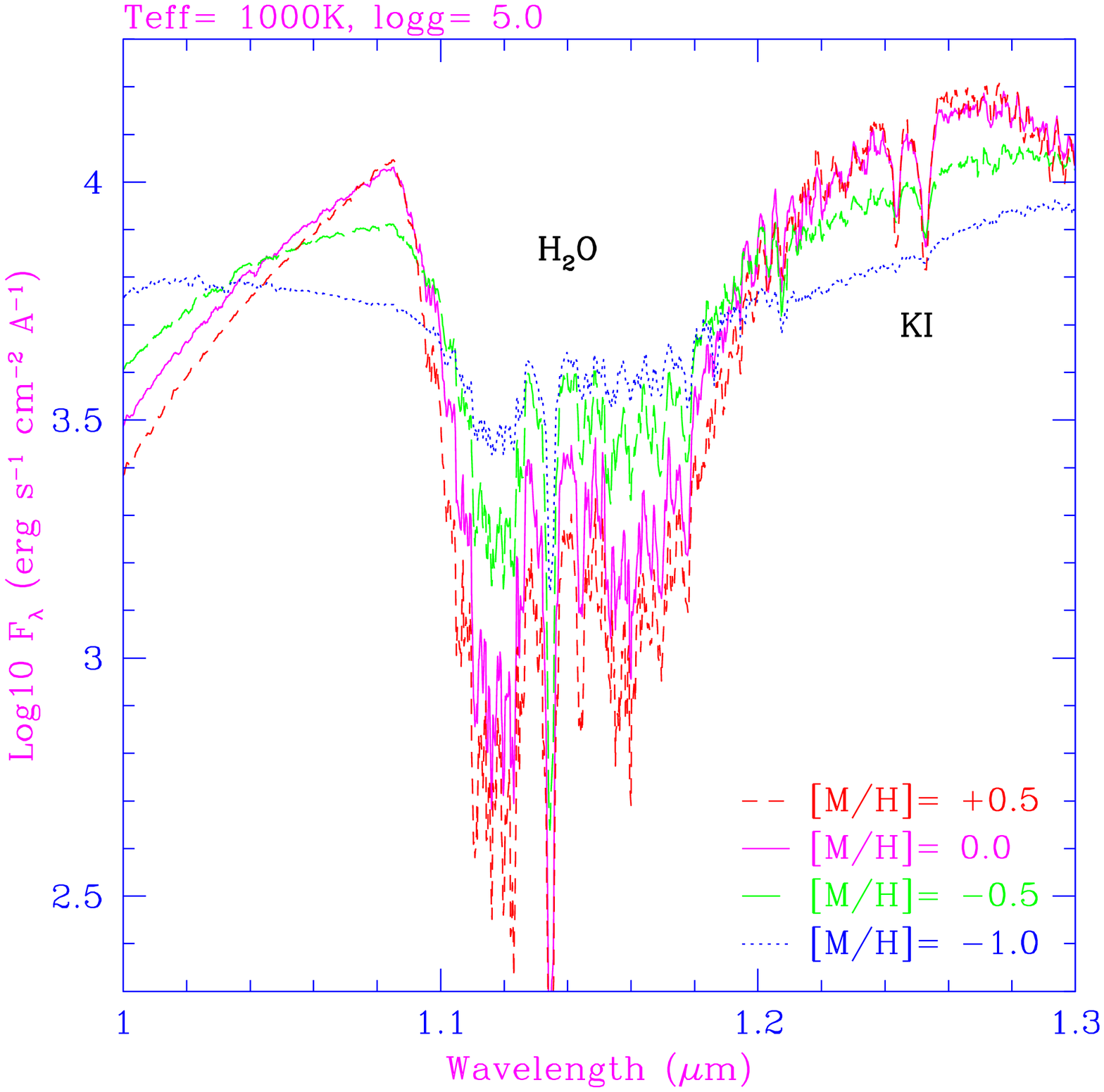,scale=0.4}
\epsfig{file=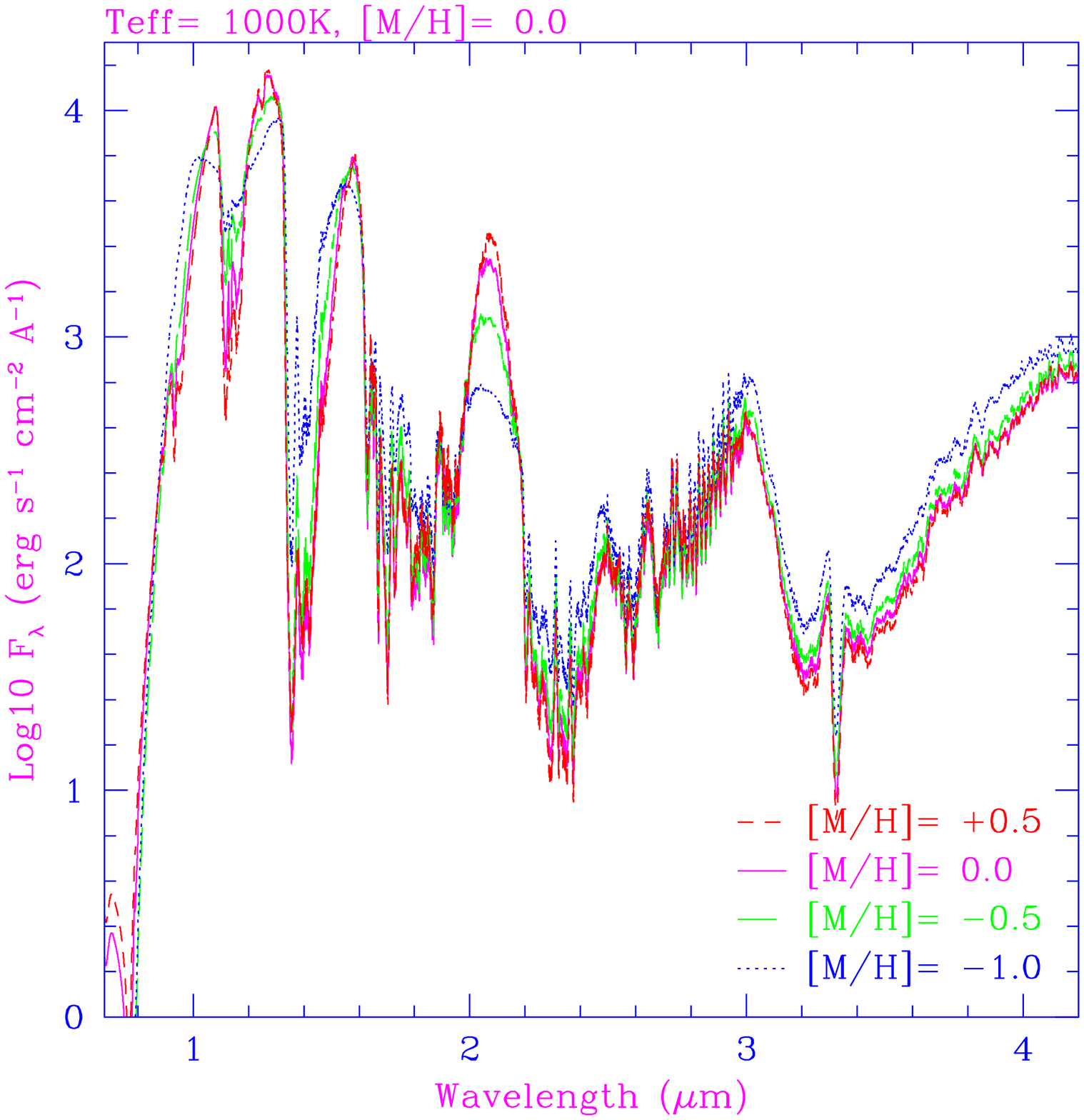,scale=0.4}
\caption{
AMES-Rainout models for $T_{\rm eff}=1000$~K, log$g=5.0$ and
[$m/H$]$=$+0.5, 0.0, -0.5 and -1.0. With decreasing 
metallicity, the broadening of the Na~ID and 0.77~$\mu$m K~I  
first increases with pressure.  This produces  
backwarming at longer wavelengths causing hotter 
H$_2$O and CH$_4$ bands characterized by weaker, broader bands 
(especially at 1.14 and 1.4~$\mu$m).  The contrast of 
the K~I doublet at 1.25~$\mu$m thus appears to decrease  with 
decreasing metallicity.  The $K$--band peak is depressed by 
increasing H$_2$ opacity.  The blue wing of the 
$Z$--band peak at 1.09~$\mu$m is determined by the K~I 
0.77~$\mu$m red wing, which for $\lambda >$ 0.88~$\mu$m
shows an unexpected decrease with decreasing metallicity most likely
due to backwarming effects.  
}
\label{default}
\end{center}
\end{figure}

Metallicity spectral signatures have only recently been explored among
newly identified L--type halo subdwarfs.  One of the first discoveries,
2MASS 0532+8246 (Burgasser et al.\ 2003), has an optical and $J$--band spectral
morphology reminiscent of a late-type L dwarf, but exhibits a highly suppressed
$K$-band peak due to enhanced H$_2$ absorption.  As a result, this
object has near--infrared colors more typical of late-type T dwarfs ($J-K \sim 0.3$).
2MASS 0532+8246 also exhibits enhanced hydride bands and broadened alkali lines,
typical characteristics of late M--type subdwarfs.  Two other L subdwarfs have
since been identified (Lepine, Rich, \& Shara 2003; Burgasser 2004).  Further discussion
on L and T subdwarfs can be found in the contribution by A.\ J.\ Burgasser (these proceedings).
No unambiguous halo (and hence metal--poor) T dwarf has yet been
identified, although the unusually blue ($J-K = -1.1$; Golimowski et al.\ 2004)
2MASS 0937+2931 (Burgasser et al.\ 2002) is a good candidate.

\subsection{Theoretical Indicators of [$m/H$], log$g$ 
({\it FA})}

Brown dwarfs, unlike main sequence stars, do not define a unique spectral sequence.
Instead luminosity is a function of age, mass and composition.  This implies
a dependency between effective temperature, surface gravity and composition.
It is important to be able to disentangle  spectral changes caused by changes in these three parameters.
In this section, we explore the response of our synthetic spectra to gravity and metallicity changes
in the two characteristic regimes of dusty and dust free atmospheres, i.e. in the M to L
type dwarf regime, and in the T dwarf regime.  We avoid the 
intermediate regime at $T_{\rm eff}$ between 1700 and 1400~K,
where partial cloud coverage and/or formation may prevail, which is discussed in \S 3.

Here we show plots demonstrating the effects of gravity (log~$g$) and
metallicity ([$m/H$]).  We use new extensions of the Dusty models grid to a wide range of metallicity, and also
a new grid of dust--free models where the grain opacities have not only
been neglected (Cond--style models), but where the chemical equilibrium has been systematically depleted of all grains (Rainout models).  These models also include new detailed
line profiles for the Na~I D and K~I doublets at 0.59 and 0.77 $\mu$m (Allard et al. 2003).


The models indicate that an increase in gravity produces similar spectral changes to
a decrease in  metallicity, since both these changes increase the gas
pressure in the atmosphere.  However, the presence of dust grains in the atmospheres of 
late M to L dwarfs, which responds strongly to a decrease in metallicity, reverses
the behavior of optical to red spectral features allowing the effects to be
separated.  This is not the
case for the dust--free T dwarfs which behave much more like M subdwarfs.     
  
The changes in the strength and width of the most prominent spectral features as a function
of surface gravity and metallicity for these two  cases are reported in
Table 1.  
Changes in the shape and brightness of the inter--band flux peaks at 1.3, 1.6 and
2.2 $\mu$m (labeled $J$, $H$ and $K$ respectively), and caused by the local 
temperature--dependance 
of the shape of the water  bands, are also reported.
Note that the width of the optical K~I and Na~I D lines and the 
pressure--induced
absorption by H$_2$ in the $K$ bandpass both increase with increasing mass density of the
plasma.  
    


\begin{table}[!h]
\caption{Summary of Effects of Gravity and Metallicity}
\label{tab:table}
\begin{center}
\leavevmode
\scriptsize
\begin{tabular}[h]{cccccccccc}
\hline \\[-5pt]
$T_{\rm eff}$ & $\rho \Uparrow$ & Lines & Grains &  TiO & H$_2$O &
H$_2$ & $J$ & $H$ & $K$ \\[+5pt]
K & & & & VO & CH$_4$ & & & & \\
\hline \\[-5pt]
2000 & $g \Uparrow$ & $\Uparrow$  & $\Uparrow$ &  $\Downarrow$ &$\Downarrow$ &$\Uparrow$ & $\Downarrow$ &$\Downarrow$ &$\Downarrow$\\
2000 & $z \Downarrow$ & $\Uparrow$  & $\Downarrow$ & $\Downarrow$ & $=$ &$\Uparrow$ & $\Uparrow$ &$\Downarrow$ &$\Downarrow$\\
1000 & $g \Uparrow$ & $\Uparrow$  & ... & ... &$\Downarrow$ &$\Uparrow$ & $=$ &$=$ &$\Downarrow$\\
1000 & $z \Downarrow$ & $\Uparrow$  & ... & ... &$\Downarrow$ &$\Uparrow$ & $\Downarrow$ &$=$ &$\Downarrow$\\
\hline \\
\end{tabular}
\end{center}
\end{table}

\subsection{Discussion ({\it All})} 

$Q$: It does seem that distinguishing between the various parameters requires
a good set of e.g. age calibrators.\break
$A$: Unfortunately most of the new L and T dwarfs are free--floating isolated objects.  But some things help --- even if metallicity and
gravity produce the same effect, the size of the effect can be very different.  Note also that if metallicity is reduced it affects the 
double--metal features (e.g. TiO, VO) more than the single--metal features 
(e.g. K~I, H$_2$O).\break
$Q$: We should ensure that any label for gravity, e.g. $abc$, provides
sufficient range to cover all expected gravities.\break
$A$: This is true.  Eventually we should use a roman numeral giving an actual value in cgs units, as for the stars.\break
$Q$: Do the speakers have favorite metallicity or gravity indicators?\break
$A$: The shape of the $ZJHK$ flux peaks are very useful, as are the K~I lines and the H$_2$O wings.  We also need to obtain a large enough sample to find the outliers with extreme metallicity and/or gravity.  One way to do this is
through proper motion surveys which can identify the older, likely metal--poor dwarfs.

\section{The Transition from L to T}

\subsection{The Sinking Homogenous Cloud 
({\it TT})}




\begin{figure}[!!b]
  \begin{center}
    \epsfig{file=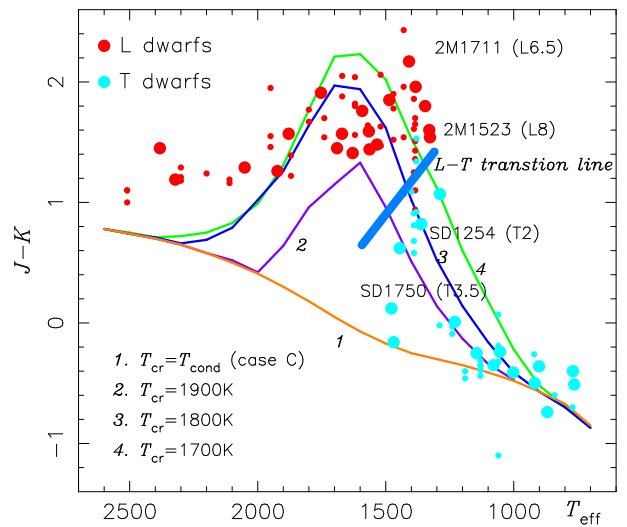, width=8cm}
  \end{center}
\caption{$J-K$ (Knapp et al. 2004) plotted against $T_{\rm eff}$ (Vrba et al. 
2004), and the predicted colors based on the UCMs with log\,$g$ =5.0 and
$T_{\rm cr} = T_{\rm cond}$ (case C), 1900, 1800, and 1700\,K.
The L---T transition line is indicated.
}
\end{figure}

In the photosphere of ultracool dwarfs, we assume that dust forms at the 
condensation temperature $T_{\rm cond}$ but soon segregates at slightly 
lower temperature which we referred to
as the critical temperature $T_{\rm cr}$. As a result,  a dust cloud 
forms in the region where $T_{\rm cr} \la T \la T_{\rm cond}$, and this model 
is also referred to as the Unified Cloudy Model (UCM, Tsuji 2002).  Since 
$T_{\rm cond} \approx 2000$\,K for iron grains, for example, the dust cloud
forms in the optically thin region of L dwarfs whose $T_{\rm eff}$'s
are relatively high, but in the optically thick region of T dwarfs
whose $T_{\rm eff}$'s are relatively low. As a result, it looks as if a 
homogeneous cloud formed in the upper photospheric layer in L dwarfs sinks to 
the deeper layer in T dwarfs. As long as the homogeneous cloud is in the
upper photosphere as in L dwarfs, it will directly effect the observables
and explains why L dwarfs appear to be dusty.  In T dwarfs, the effect of dust on
the observables diminishes  as the dust cloud sinks to
the deeper region and, to a first approximation, this immersion
of the homogeneous cloud explains the transition from L to T.
Note, however, that no specific mechanism is assumed for the 
sinking of the homogeneous cloud, this is simply a natural consequence  of the 
change of the thermal structure as L dwarfs evolve to T dwarfs.


This is clearly shown in Figure 9 where  $J-K$ (Knapp et al. 2004) 
is plotted against  $T_{\rm eff}$ based on the bolometric flux
(Leggett et al. 2002, Golimowski et al. 2004, Vrba et al. 2004), 
with different symbols for L and T dwarfs. The predicted values of $J-K$
for several values of $T_{\rm cr}$  are overlaid. The lower
value of $T_{\rm cr}$ implies that the homogeneous cloud is thicker
and $J-K$ will be redder because of the increased dust extinction.
The scatter of the observed $J-K$ at any $T_{\rm eff}$ is rather large 
and this means that $T_{\rm cr}$ is variable at a fixed $T_{\rm eff}$. 
After all, both $T_{\rm eff}$ and  $T_{\rm cr}$ change in the transition
from L to T across the ``L---T transition line'' indicated by the thick 
line in Figure 9.


\begin{figure}[!!t]
  \begin{center}
    \epsfig{file=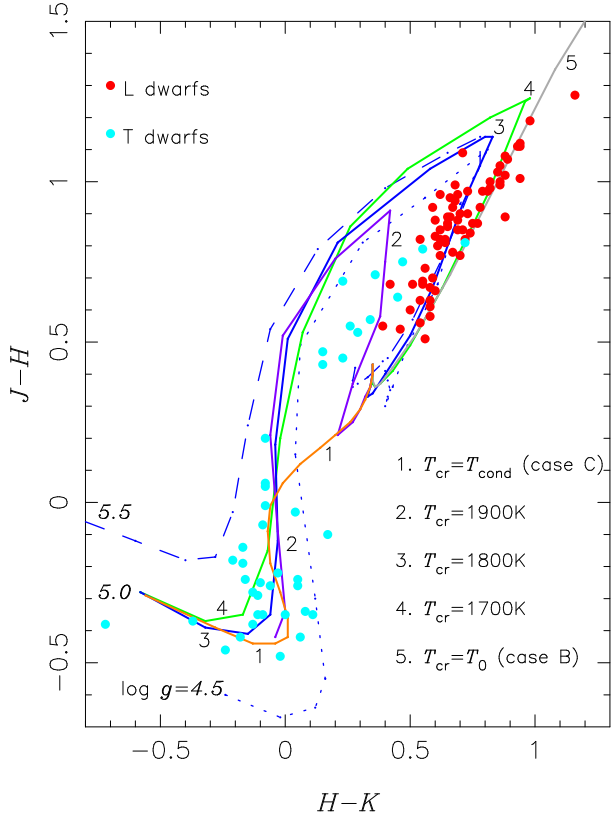, width=8cm}
  \end{center}
\caption{Observed $(J-H, H-K)$  (Knapp et al. 2004) compared with
the predicted colors based on the UCMs with  $T_{\rm cr} = 
T_{\rm cond}$ (case C), 1900, 1800, 1700\,K, and $T_{\rm 0}$ (case B) under
the fixed value of log\,$g$ =5.0, and those for log\,$g$ =4.5 (dotted line)
and 5.5 (dashed line) with the fixed value of $T_{\rm cr} = 1800$\,K.
Note that $700 \leq T_{\rm eff} \leq 2600$\,K throughout.
}
\end{figure}

In Figure 10, the observed  $(J-H, H-K)$ (MKO system, Knapp et al.
2004) is compared with the predicted values for several values of $T_{\rm cr}$
at the fixed value of log\,$g$ = 5.0. 
The cases of log\,$g$ = 4.5 and 5.5 are shown by the dotted and dashed 
lines, respectively, at the fixed value of $T_{\rm cr} = 1800$\,K. Inspection
of Figure 10 reveals that the spread of the observed data in L and early T 
dwarfs 
is explained by the continuous change of $T_{\rm cr}$ from $T_{\rm cond}$ to
$T_{\rm 0}$ (surface temperature) while that in  late T dwarfs by the effect 
of log\,$g$. Thus, observed two--color diagrams can  be explained with the three 
parameters, $T_{\rm eff}$, log\,$g$, and $T_{\rm cr}$, but cannot with 
the two parameters, $T_{\rm eff}$ and  log\,$g$ alone.


\begin{figure}[!!t]
  \begin{center}
    \epsfig{file=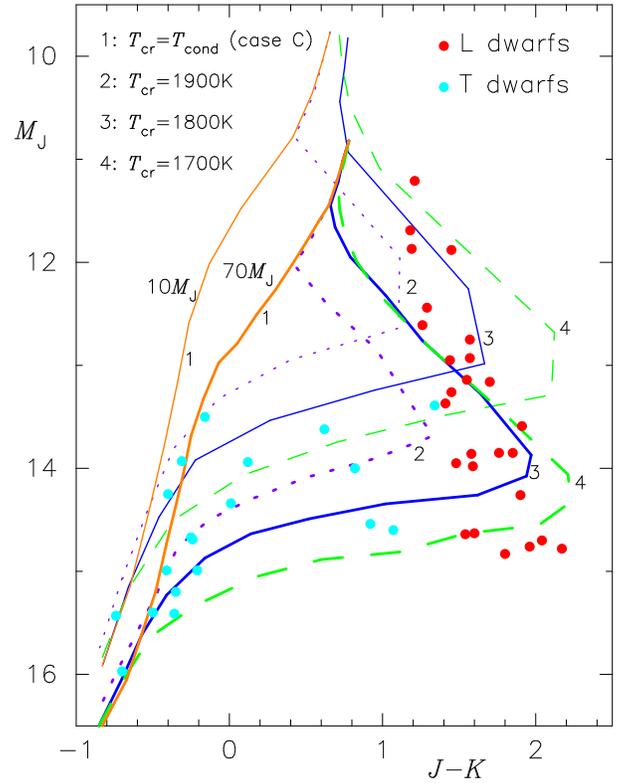, width=8cm}
  \end{center}
\caption{Observed  $(J-K, M_J)$ diagram (Knapp et al. 2004)
compared with predicted loci which are transformed from 
evolutionary tracks (Burrows et al. 1997) of $M/M_{\rm Jupiter}$ = 
10\,(thin lines) and 70\,(thick lines), via the UCMs 
with $T_{\rm cr} = T_{\rm cond}$ (case C), 1900, 1800 and 1700\,K.
}
\end{figure}

Characteristic features of the observed $(J-K, M_J)$ 
color--magnitude (CM) diagram 
(Dahn et al. 2002, Tinney, Burgasser \& Kirkpatrick 2003, Vrba et al. 2004) 
shown in Figure 11 are 
rapid bluing and brightening at the transition from L to T.  
We transform the theoretical ($T_{\rm eff}, M_{\rm bol}$) diagrams 
for initial masses of 10 and 70\,$M_{\rm Jupiter}$ (Burrows et al. 1997)
to the $(J-K, M_J)$ diagram via  UCMs with $T_{\rm cr} = T_{\rm cond}$, 
1900, 1800, and 1700\,K, instead of the previous attempt assuming a uniform 
value of $T_{\rm cr} = 1800$\,K throughout (Tsuji \& Nakajima 2003), and 
the results are overlaid on Figure 11. It is found that almost all the
observed data can be reproduced with our predictions, and the bluing and
brightening of the early T dwarfs can be explained by the models of
$T_{\rm cr} \approx T_{\rm cond}$ (i.e. effectively no cloud) if not by the 
very low--mass models. Thus the L---T transition on the CM diagram can be 
explained with the sinking homogeneous cloud model, but only if a sporadic  
variation of $T_{\rm cr}$, which is a measure of the thickness of the cloud 
(or dust column density) in the observable photosphere, is assumed.
Such a variation of $T_{\rm cr}$ is not predicted by the present theory of 
structure and evolution of substellar objects, but we had to introduce 
$T_{\rm cr}$ as a free parameter to interpret purely empirical data
such as the two--color diagram and CM diagram.


\begin{figure}[!!b]
  \begin{center}
    \epsfig{file=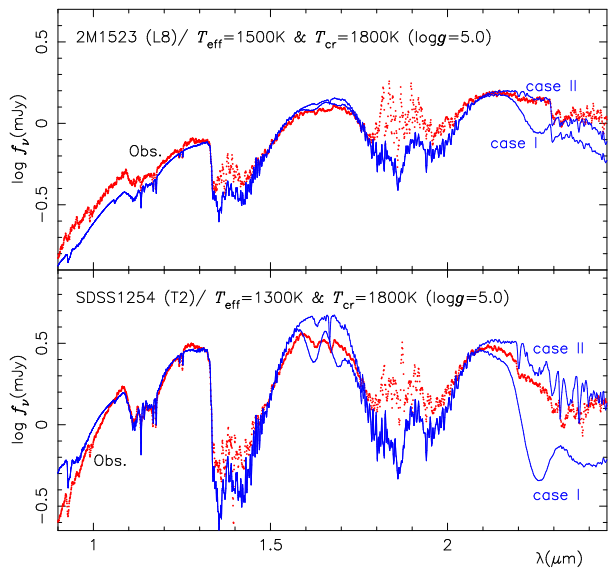, width=8cm}
  \end{center}
\caption{
The spectral change at the transition from L to T, exemplified
by 2MASS\,1523 (L8) and SDSS\,1254 (T2), can be interpreted
as due to the change of $T_{\rm eff}$ under the fixed value of $T_{\rm cr} 
= 1800$\,K. Observed and predicted spectra are shown by dots and 
solid lines, respectively.}
\end{figure}

\begin{figure}[!!b]
  \begin{center}
    \epsfig{file=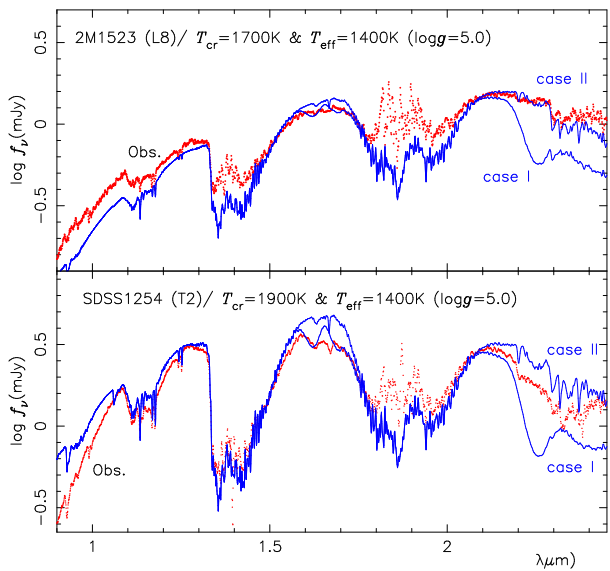, width=8cm}
  \end{center}
\caption{The spectral change  at the transition from L to T, exemplified
by 2MASS\,1523 (L8) and SDSS\,1254 (T2), can also  be 
interpreted as due to the change of $T_{\rm cr}$ under the fixed value of 
$T_{\rm eff} = 1400$\,K. Observed and predicted spectra are shown by  
dots and solid lines, respectively.
}\end{figure}

The change of the spectra at the transition from L to T could
have been interpreted as due to the change of $T_{\rm eff}$ on the
assumption of $T_{\rm cr} = 1800$\,K throughout (Tsuji, Nakajima \& Yanagisawa 2004). 
For example, the typical  L dwarf 2MASS\,1523 showing red colors and weak 
or no CH$_{4}$ bands could be consistent with $T_{\rm eff} = 1500$\,K, while 
the early T dwarf SDSS\,1254 showing strong H$_{2}$O and modest CH$_{4}$ 
bands with $T_{\rm eff} = 1300$\,K (Figure 12). However, infrared colors 
suggest 
that $T_{\rm cr}$ cannot be the same throughout L---T dwarfs (Figure 9) 
and 
also empirical values of $T_{\rm eff}$ show a plateau at $T_{\rm eff} 
\approx 1400$\,K between L8 and T4 (Golimowski et al. 2004, Vrba et al. 
2004). 
For this reason, we 
examine if the same spectra at the L---T transition can be explained with 
$T_{\rm eff} \approx 1400$\,K throughout
but with changing  $T_{\rm cr}$. The L8 dwarf 2MASS\,1523 can be explained 
with $T_{\rm cr} = 1700$\,K while the T2 dwarf SDSS\,1254 with $T_{\rm cr} 
= 1900$\,K (Figure 13). Thus, different combinations of $T_{\rm eff}$ and 
$T_{\rm cr}$ could explain the change of spectra at the transition from L 
to T. This is because the dust column density in the observable photosphere 
depends not only on $T_{\rm eff}$ but also on $T_{\rm cr}$.

How can we decide which case is correct? For this purpose,
assume first the empirical $T_{\rm eff}$ based on the bolometric flux,
and $T_{\rm cr}$ can be estimated based on the infrared colors (Figure 9). 
Also,
transform the observed spectrum to the spectral energy distribution 
(SED) on an absolute scale
with the use of the observed parallax. Then, analyze the SED to improve  
$T_{\rm eff}$, $T_{\rm cr}$,  log\,$g$, chemical composition, 
$v_{\rm micro}$, etc. Such an analysis applied to a sequence of L---T dwarfs  
including 2MASS1711 (L6.5), 2MASS\,1523 (L8), SDSS\,1254 (T2), and 
SDSS\,1750 (T3.5), which could have been interpreted as a sequence of 
$T_{\rm eff}$ from 1800 to 1100\,K with the uniform value of $T_{\rm cr} 
= 1800$\,K (Tsuji, Nakajima \& Yanagisawa 2004), revealed that $T_{\rm eff} 
\approx 1300$\,K
throughout but $T_{\rm cr}$ should change from 1700\,K to $T_{\rm cond}$
(Tsuji, these proceedings). It is surprising that the distinct changes of
the spectra from L6.5 to T3.5, including the L---T transition, are nothing to 
do with $T_{\rm eff}$ but are instead due to the change of the dust column density at 
fixed $T_{\rm eff}$. This result that the L---T spectral sequence is not
a temperature sequence is based on a limited sample, but the same conclusion 
can be suggested by the curious brightening of $M_J$ when plotted against 
the L---T types.


The transition from L to T takes place at $T_{\rm eff} \approx 1400 \pm 
100$\,K (Figures 2, 9), and this transition can be understood as a 
consequence
of a homogeneous cloud effectively sinking from optically thin to
thick regions at about this $T_{\rm eff}$.  However, the L---T transition 
appears to be a somewhat more complicated phenomenon in that $T_{\rm cr}$ also 
changes (Figure 9) and this is even more directly related to the L---T 
spectral 
types around the transition. In other words,  the thickness 
of the homogeneous cloud appears to decrease and even disappear at the transition 
from L to T. However,  the mechanism of  how $T_{\rm cr}$ changes 
at $T_{\rm eff} \approx 1400$\,K is unknown. The only known change of the 
photospheric structure is the formation of the second (surface) convective
zone at $T_{\rm eff} \approx 1400$\,K in the UCMs (Tsuji 2002).
The exact meaning of the transition from L to T remains
unsolved until the origin of the sporadic variation of $T_{\rm cr}$ can be
identified. Moreover, the meaning of the L---T spectral classification
as a whole should be reconsidered in view of the difficulty of
interpreting it as a temperature sequence.


\subsection{Other Models for the Transition 
({\it MSM})}


A sufficient number of late L to early to mid T dwarfs (approximately L8 to T5) have now been found that a number of characteristics of the transition can be listed.
\begin{itemize}
\item
{\it Turn to the blue in $J-K$}  The colors of L dwarfs become 
progressively redder until they saturate at $J-K \sim 2$ at spectral type L8 (Knapp et al. 2004).  This color then rapidly turns to the blue, reaching $J-K\sim -0.8$ by T8 or so. 
\item
{\it Color change at near constant $T_{\rm eff}$}  Recent estimates of the bolometric $T_{\rm eff}$ from Golimowski et al.  (2004, made possible by the parallax measurements of Vrba et al. 2004) have quantified the speed of this color change, as shown in Figures 
9 and 14. Most ($>80\%$) of the change in $J-K$ color is seen to occur 
over a very small 
$T_{\rm eff}$ range near 1300 K.   This is a remarkable result as it implies that brown dwarfs are undergoing substantial spectral and color changes over a very small temperature range.
\item
{\it Brightening at $J$} The L to T transition also appears to be associated with a brightening at $J$ from late L to early T (T4 or so, Knapp et al. 2004).  $H$, $K$, $L$, and $M$ bands show no sign of such brightening (Knapp et al. 2004, Golimowski et al.  2004), while there is some evidence of a brightening at $Z$.  It should be noted that the bolometric luminosity, as would be expected, does {\it not} increase across the transition.
\item
{\it Resurgence of FeH} Burgasser et al. (2002) argue there is evidence that the $0.997\,\rm\mu m$ FeH band, after decaying away as FeH is presumably lost to Fe drops and grains,  shows a resurgence in strength, coincident with the $J-K$ color change.
\item
{\it Model spectral fits}  The comparison of models and data as shown by Tsuji in the previous section and by Marley et al. elsewhere in these proceedings shows that while cloudy models fit the L dwarfs, by spectral type
 T5 models with no cloud opacity (but with condensation included in the equilibrium chemistry) fit spectra very well, implying that condensates play a very small role in controlling the thermal profile and emitted flux of mid-- to late--T dwarfs. 
\end{itemize}

Any explanation of the L to T transition mechanism must be consistent with the evidence summarized above. The unmistakable gross explanation -- that condensates have been lost from the atmosphere -- belies the difficulty in explaining this loss in a 
self--consistent manner.  That a sinking, finite--thickness cloud deck 
will eventually disappear from sight allowing the atmosphere above to cool 
has been apparent for some time (Allard et al 2001, Marley 2000, Tsuji \& Nakjima 2003).  The difficulty lies in explaining the rapidity of the color change in light of the measured effective temperatures (Figures 9 and 14).  The cloud model of Ackerman \& Marley (2001) while nicely accounting for the $J-K$ colors of the reddest L dwarfs takes much too long to ultimately sink out of sight (Burgasser et al. 2003, Knapp et al. 2004).  

Tsuji (2002) and Tsuji et al. (2004)  proposed that a physically thin cloud, thinner than predicted by the 
Ackerman \& Marley model, could self--consistently explain the rapid L to T transition.  These UCM models indeed 
exhibit a faster L-- to T--like transition, but as Figure 9 demonstrates these models, with fixed 
$T_{\rm cr}$, are not consistent with the observed rapidity of the color change.  Even accounting for a likely 
spread in gravities across the transition cannot account for the observations. In addition the 
UCM models, like the cloudy models of Marley et al., do not brighten in $J$ band across the transition.  
Tsuji now suggests (\S 3.1) that $T_{\rm cr}$ must vary across the transition, allowing the cloud to 
essentially collapse as $T_{\rm cr} \rightarrow T_{\rm cond}$.

\begin{figure}[!!b]
\begin{center}  
\includegraphics[height=.4\textheight,angle=-90]{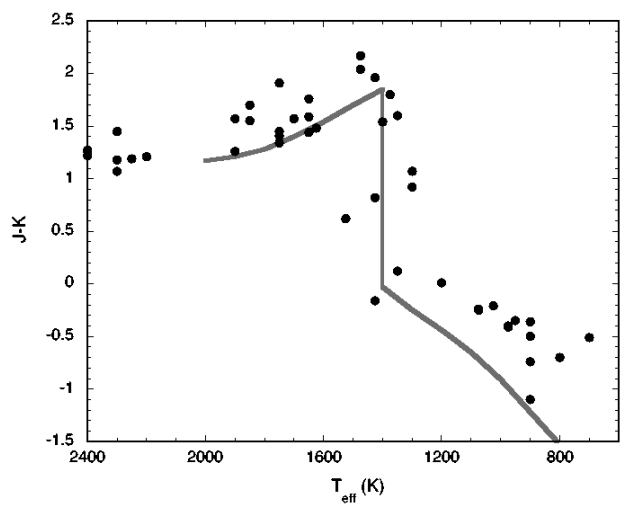}
\end{center}
\caption{ Ultracool dwarf MKO $J-K$ and $T_{\rm eff}$ (Knapp et al. 2004 and
Golimowski et al. 2004).  Model curve is a composite of cloudy (for $T_{\rm eff} \ge 1400\,\rm K$)
and clear (for  $T_{\rm eff} \le 1400\,\rm K$) models (Marley et al. 2002) connected by a vertical line at $T_{\rm eff}=1400\,\rm K$, all for $\log g = 5$.} 
\end{figure}

To overcome the sort of difficulties faced by the Tsuji et al. models, Burgasser et al. (2002), following a suggestion from
Ackerman \& Marley (2001),  hypothesized that the transition was 
associated with the appearance of holes in the global cloud deck.  The 
holes, hypothesized to be similar to Jupiter's well known ``5-micron hot spots'', would allow flux to emerge through the $Z$-- and $J$--band atmospheric opacity windows.  If the onset of the holes corresponded with the $T_{\rm eff}$ of the L/T transition region, then their appearance could account for the characteristics of the transition outlined above.  Indeed Burgasser et al. demonstrated with a simple model that holes could plausibly account for the bluing in $J-K$ and brightening in $J$.  Figure 14, similar to Figure 9, demonstrates that, for fixed gravity, rapidly moving from a cloudy to a cloud--free
model apparently accounts for the available data.  This mechanism would also account for the reappearance of FeH absorption since hot, FeH--bearing gas would be detectable through the clouds, if the holes pierced both the global silicate and underlying Fe cloud deck.

Finally Knapp et al. (2004) suggested that, like a varying $T_{\rm cr}$, a varying $f_{\rm sed}$ (sedimentation efficiency, Ackerman \& Marley 2001) could result in a sudden downpour that might rapidly remove the cloud deck over a small effective temperature range.

Regardless of whether the L to T transition is explained by the appearance of 
holes in the global cloud deck or a sudden increase in the efficiency of 
condensate sedimentation, the root cause must lie with the atmospheric 
dynamics.   Perhaps the behavior of condensates change when the cloud reaches 
a certain depth in the atmospheric convection zone.  When even the cloud tops 
are firmly rooted in the convection zone the clouds may be sheared apart by 
height--varying zonal flow, like that found in Jupiter. Perhaps the second, 
detached convection zone found in brown dwarf atmosphere models plays a role.  
Another possibility is that there is a change in the global atmospheric 
circulation that affects the cloud decks.  Schubert \& Zhang (2000) found that 
brown dwarfs likely exhibit one of two styles of global atmospheric 
circulation.  They may either exhibit circulation strongly influenced by rotation, 
like Jupiter, or fairly independent of rotation, like the Sun.    
Perhaps the L to T transition is associated with a transition between
the two circulation regimes encountered as a brown dwarf cools.
The solution of this enigma 
certainly lies in understanding the three dimensional interplay of global 
atmospheric circulation with both macro-- and micro--scale cloud dynamics.

\subsection{Discussion ({\it All})}

$Q$: We need to see the L to T transition in clusters to reduce the
age spread in the color:magnitude diagram.\break
$A$: Work done on the $\sigma$ Ori cluster is probing into the T
regime and may show the blueward turn in the diagrams.  \break
$Q$: The change in spectral type at constant effective temperature is
a real and observed efect; it is not artificial.
\break
$A$: Its artifical only in the sense that it is due to changing cloud parameters  
and not changing $T_{\rm eff}$. \break
$Q$: Is it OK to have spectral type $not$ depend
on $T_{\rm eff}$?\break
$A$: In the spirit of the spectral type being an observable, yes it is.  The
spectra definitely change.\break
$Q$: Is there a correlation between $T_{\rm cr}$ and $f_{\rm sed}$?\break
$A$: No because there are differences between the two models such as particle size.  Note that no set of parameters for particle size can explain the L to T transition.\break
$Q$:  Surely microphysical effects should be included?\break
$A$: This is very complex and we try to mimic the effects through the 
$T_{\rm cr}$ and $f_{\rm sed}$ parameters.  [Discussion of the microphysics is presented by Hellings elsewhere.]\break
$Q$: Is the mixing length approximation to convection valid?\break
$A$: It is acceptable to first order for L and T dwarfs.  The plane parallel assumption is OK too.\break
$Q$: For Jupiter we see that the holes are significant flux contributers.\break
$A$: The L and T dwarfs are being monitored for variability.  Interpreting the results depends however on both hole size and number.

\section{Spectral Types Beyond T  
({\it HRAJ})}

\begin{figure}[!!b]
  \begin{center}
\includegraphics[height=.35\textheight,angle=-90]{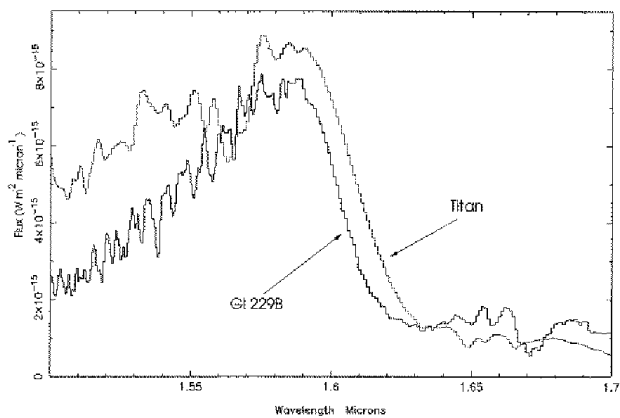}
  \end{center}
\caption{Spectra of Gl~229B (solid line) and Titan
(scaled by a factor of 0.003) near the 1.6$\mu$m methane 
absorption edge (credit - Geballe et al. 1996).\label{jones_geballe}}
\end{figure}

The concepts and nomenclature of stellar classification were refined in
the late 19th century primarily through the works of Secchi, Fleming and
Pickering. The sequence of letters used today comes from the work of Canon
\& Pickering (1901) who produced the OBAFGKM empirical sequence based on
differences between photographic spectra. This sequence has been
embellished with further spectral types, e.g. C to denote carbon stars
(Morgan, Keenan \& Kellman 1941), but served as a complete temperature
sequence for nearly a hundred years.  Then in the 1990s intrinsically
faint very cool objects started to be discovered (e.g. GJ 229B by Nakajima et al.
1995). The spectra of objects such as GJ 229B (e.g. Oppenheimer et
al. 1995 and Figure 15) showed unprecedented methane--rich spectra. Such 
spectra along with the burgeoning numbers of M dwarfs with spectral types 
beyond M9 (Kirkpatrick et al. 1998) gave rise to the necessity for further 
spectral types.


The letter L results from a proposal by Mart{\'{\i}}n et al. (1997) that L 
should denote Low temperature and there might be Low temperature lithium and
methane designations. The L spectral type and also the T spectral type
became generally accepted in the literature following a seminal 
paper by Kirkpatrick et al. (1999) and a meeting dedicated
to the ``Ultracool Dwarfs: New Spectral Types L and T'' (Jones \& Steele
2001). Despite these relatively recent introduction of new
spectral classifications, even more recent discoveries of very faint red 
objects, supported by infrared spectra and parallaxes (Burgasser 
et al. 2002, Geballe et al. 2002, Golimowski et al. 
2004, Knapp et al. 2004, Tinney et al. 2003, Vrba et al. 2004) indicate 
that a spectral type 
of T8 or T9 has been reached.  It is thus important to consider further 
spectral types.

\begin{figure}[!!b]
  \begin{center}
    \epsfig{file=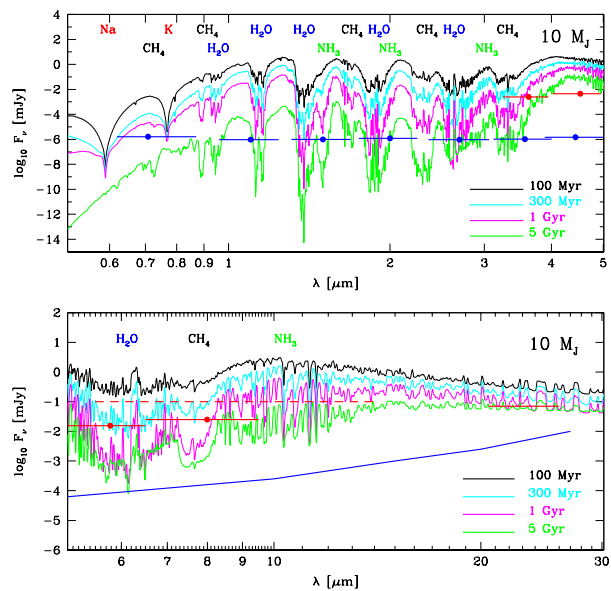, width=8.5cm}
  \end{center}
\caption{Spectra of 10 M$_{\rm Jup}$ brown dwarfs at
1, 3 and 5 billion years. the $SIRTF$ and $JWST$ detector sensitivities
are plotted for reference. Gaseous water absorption features remain
strong over this range of ages. Methane and ammonia absorption features
strengthen with age, as the
alkali lines wane. The wavelength positions of various of the molecular
and atomic features are depicted for reference. (credit - Burrows et al.
2003).\label{jones_burrows}} \end{figure}

\begin{figure}[!!b]
  \begin{center}
\includegraphics[height=.35\textheight,angle=-90]{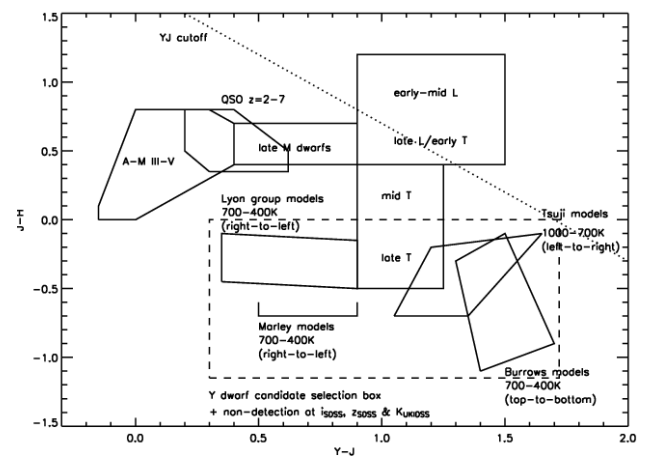}
  \end{center}
\caption{The $J-H$, $Y-J$ two--colour diagram for sources
we expect in the WFCAM LAS survey. The location of 
A--M (III--V) stars, late M, L and T dwarfs and z$=$2---7 
QSOs have been determined using colours synthesized 
from available spectroscopy. Y dwarf
colours have been synthesized from the latest
theoretical model spectra by the Lyon group (priv.
comm.), the Marley group (priv. comm.; see Marley et 
al. 2002), the Burrows group
(Burrows et al., 2003) and Tsuji
(priv. comm.). The Lyon $J-H$ colours have been
adjusted by $-0.3$ so that they pass through the
known T dwarfs. \label{jones_pinfield}} \end{figure}

Notwithstanding the non--alphabetical order of the hot spectral types, it 
seems desirable to continue the sequence
alphabetically when suggesting further, cooler, types. Thus
a letter from T to Z would be preferred, though perhaps not Z itself as 
this suggests the end of the spectral typing sequence. There are three 
major factors to
consider in choosing a new spectral type: (1) it must be unambiguous and
should not currently be used to represent any other spectral type, (2) the
letter must represent a typing that is clearly distinguished from other
types of astronomical objects, (3) the letter should be free of
physical interpretation (which is likely to vary with time). As discussed
by Kirkpatrick et al. (1999), U and X are problematic choices because of
possible associations with Ultraviolet and Xray sources. V would be a good
choice but for the possible confusion with
vanadium oxide (VO) which shows prominent bands in spectra of
late--type M and early--type L dwarfs. W is undesirable
because of likely confusion with Wolf-Rayet WN and WR classes.
Y could be confused with yttrium oxide (YO) which may also appear in cool
dwarf spectra, however, it has yet to be identified and the low abundance 
of YO means that it may not be evident. Assuming that YO is not found, 
Y seems the best choice for the next spectral class.
While the choice of Y may seem uncomfortably close to the end of the
alphabet it is important to note that the broad spectral features of T
dwarfs are somewhat similar to the solar system planets Titan and Jupiter.
Although the infrared spectra of solar
system objects are in reflected sunlight, their spectra are dominated by
methane and water vapour absorption features in a similar manner to T
dwarfs, e.g. Figure \ref{jones_geballe}. In fact, given the relatively 
low masses and temperatures of Titan and Jupiter it is perhaps desirable 
that the classification system is reaching an alphabetic end point.

The dominance of methane in the infrared spectra of T dwarfs means they
are often called methane dwarfs. While methane is the defining
characteristic of T dwarfs it can also be seen at 3.3 $\mu$m in L spectral 
types later than L4   (Noll et al. 2000).  Recently obtained 
mid--infrared spectra (Roellig et al. 2004) show that ammonia can be seen 
at 11~$\mu$m in T dwarfs later than T5.
Figure \ref{jones_burrows} from Burrows, Sudarsky \& Lunine (2003)
indicates that ammonia will become dominant in both the near-- and 
mid--infrared  at temperatures below 700~K.  It is likely that, in 
the same manner that T dwarfs are known as methane dwarfs, 
the next spectral type will be known as ammonia dwarfs. 

Figure \ref{jones_pinfield} shows a colour--colour plot with 
model predictions for Y dwarfs, prepared to aid selecting such
objects in the UK Infrared Telescope's (UKIRT) Wide Field Camera (WFCAM)
Large Area Survey (LAS).  Although
the model groups predict somewhat different colors, we can identify candidate
Y dwarfs as red in $Y-J$ and blue in $J-H$ (see also the WFCAM contribution 
by Leggett in these procedings).  A detailed
understanding of Y dwarf properties (and ultimately the brown
dwarf mass function) will require spectroscopic analyses to
simultaneously determine effective temperature, metallicity and gravity,
as we are now attempting for L and T dwarfs.


\subsection{Discussion ({\it All})}
%

A short contribution on characterization of exoplanetary atmospheres with
the VLT Interferometer was presented, see Joergens \& Quirrenbach, these
proceedings.

\end{document}